# Active Collaborative Filtering

Craig Boutilier and Richard S. Zemel and Benjamin Marlin
Department of Computer Science
University of Toronto
Toronto, ON, M5S 3H5, CANADA
cebly,zemel,marlin @cs.toronto.edu

## Abstract

Collaborative filtering (CF) allows the preferences of multiple users to be pooled to make recommendations regarding unseen products. We consider in this paper the problem of online and interactive CF: given the current ratings associated with a user, what queries (new ratings) would most improve the quality of the recommendations made? We cast this in terms of *expected value of information (EVOI)*; but the online computational cost of computing optimal queries is prohibitive. We show how offline prototyping and computation of bounds on EVOI can be used to dramatically reduce the required online computation. The framework we develop is general, but we focus on derivations and empirical study in the specific case of the multiple-cause vector quantization model.

## 1 Introduction

Collaborative filtering (CF) has attracted considerable attention over the past decade due to the ease of online data accumulation and the pressing need in many applications to make suggestions or recommendations to users about items, services or information. When other users have viewed a product of interest and offered ratings of that product, the existing ratings can be used to predict the rating of a subject who has not seen the product. Specifically, if users with similar "interests" to the subject (as determined using ratings by the subject on other products) have rated the product in a particular way (e.g., positively), we might predict that the subject would rate the product in the same way (e.g., recommend the product). In this way, CF allows the preferences of multiple users to be pooled in a principled way in order to make recommendations. The CF approach forms the basis of many recommender systems [2, 6, 7, 5, 4], applied to areas as diverse as books, movies, jokes, and newsgroup articles.

A number of different approaches to CF have been proposed, including correlation analysis [6], naive Bayes classifiers [2], latent class models [5], PCA [4], and vector quantization [10]. Many of these construct explicit probabilistic models of the domain, positing features or clusters of users and/or products, and relating user and product features to predicted ratings. It is natural to ask in such settings whether additional user ratings can increase the quality of recommendations made for that user. If a user has rated   products, and we can ask the user to rate a   product, we want to know: (a) whether this new rating can improve our predictions or the value of the recommendation we make; and (b) which product offers the greatest expected benefit in this regard. An *active* approach to CF involves asking queries of this type when the expected benefit outweighs the cost (e.g., delay, bandwidth, or cognitive burden) associated with the query.

Approaches to CF that learn explicit probabilistic models of the domain facilitate the analysis of this problem: we can pose it in terms of *expected value of information (EVOI)*. Assuming some measure of utility associated with recommendations, we can compare the expected value of the best recommendation w.r.t. our current ratings distribution for a user with the expected value of the (typically different) optimal recommendation for the posterior obtained by updating with the response to that query. This form of *myopic* EVOI can be used to select queries. This model can be especially useful when dealing with users who have not yet populated rating space sufficiently (particularly new users), and allows maximum benefit to be derived from fewer product ratings. It is also useful in cases where we have low confidence in predicted ratings. In such settings, the benefit of having a user rate several additional unseen products (e.g., by playing a music or movie clip) may be substantial.

Unfortunately, computing (myopic) EVOI exactly is computationally difficult. In principle, we could ask a user about any unrated product, and for each possible response (i.e., rating of that product), we must generally compute the posterior over the ratings of the remaining prducts to determine the new optimal decision. Worse yet, this computation must be performed online, while interacting with the user. Since CF is most useful in situations with large numbers of users and products, this in unlikely to be feasible except in the most trivial settings.

We consider approaches that allow us to bound the expected changes in these posteriors in a user-independent fashion. By constructing such bounds offline (using the learned model), we can dramatically reduce the number of online posterior computations needed to determine the query with maximum EVOI. In addition, we can use properties of the learned model to construct a small set of *prototype* users and queries, further reducing the online computational complexity, with only a small sacrifice in decision quality. The framework we develop is quite generic, and can be applied to any CF algorithm that produces an explicit probabilistic model of the domain. However, the details will depend on the specifics of the model in question. We develop these details for the specific case of the *multiple-cause vector quantization (MCVQ)* model developed by Ross and Zemel [10]. However, the development will be similar for most other types of probabilistic models (e.g., naive Bayes, general Bayesian network etc.) commonly used for CF. We will demonstrate our approach with a naive Bayes model to emphasize this point.

Active CF has been suggested by Pennock and Horvitz [8]; but their work does not explore such methods, nor does it suggest techniques for implementing the active component in the face of the intensive online computational challenges associated with the use of EVOI. Active querying has recently been explored in the context of the "new user problem" [9]. This work, however, suggests simple heuristic measures for populating a user database with ratings that do not rely on specific utility-theoretic measures. Indeed, our work is distinguished from most existing CF research in its focus on the *value* of recommendations, rather than overall predicted rating accuracy. This is key to the success of EVOI: one need not worry about accuracy of predicted ratings that have no impact on the decision or recommendations one will make. Our work is also related to more generic forms of active learning (e.g., [3]), though our focus is on the more specific details of CF and ensuring that online computation is tractable.

The remainder of the paper is organized as follows. In Section 2, we discuss CF, and introduce the notion of EVOI for active collaborative filtering in the context of generic probabilistic models. In Section 3, we spell out the details of the active framework in the specific case of the MCVQ model. We show some simple experiments illustrating the benefit of EVOI-based active querying in both the MCVQ and a naive Bayes model. Section 4 details a method for bounding the impact a query can have on the mean rating of a target product in a user-independent (offline) fashion, allowing the query with maximum EVOI to be computed more effectively online. Empirical results demonstrate that a significant amount of pruning can be obtained in the MCVQ model. We discuss some preliminary ideas pertaining to offline prototyping of queries in Section 5, which further reduces the space of queries one needs to consider and again show the benefits empirically. We conclude with further discussion of related work, some suggestions for refinements to the model and directions for future research.

## 2 A Framework for Active CF

We begin by establishing notation and basic background on CF. We then define EVOI in the context of generic probabilistic CF models, and describe an active approach to CF.

### 2.1 The Collaborative Filtering Problem

The basic task in CF is to predict the utility of items from some set $M$ to the target or *current* user based on a database of ratings from a population of other users. We assume ratings are provided explicitly, on a given scale. From a probabilistic perspective, the aim is to estimate the probability that the current user will assign a particular rating to an as-yet unobserved item. The basic paradigm in CF, especially in learning-based models [2, 5, 10], the training set of user ratings produces model parameter values, which permit the online estimation of these probabilities based on the set of items for which the active user has provided ratings. Batches of user data can also be used to update the parameter values.

Let $P$ denote the distribution over rating vectors for a generic CF model, trained on existing data. Let $\kappa_i$ denote the set of products for which user $i$ has provided ratings; since the user $i$ will generally be fixed throughout our discussions, we typically drop the subscript. Let $\mathbf{r}_\kappa$ denote the vector of ratings over this set $\kappa$ and $\overline{\kappa} = M \setminus \kappa$. Given the current ratings vector $\mathbf{r}_\kappa$, we obtain a posterior distribution for each $j \in \overline{\kappa}$:

$$P_{\mathbf{r}_\kappa}(R_{ij}) = P(R_{ij}|\mathbf{r}_\kappa) \quad (1)$$

$R_{ij}$ can be treated as either a discrete or continuous variable.

Original statistical CF approaches predicted unobserved ratings using a weighted linear combinations of other users' ratings, with weights derived from the correlation between each user and the active user [6]. Latent factor models have also been applied to this problem. For example, simple naive Bayes models have been used with some success [2], as has a different form of latent factor model known as the aspect model [5]. Space precludes a detailed discussion of these models. The details of any active approach to CF will depend on the underlying probabilistic model; for this reason, we describe a specific model (the MCVQ model) in detail below. However, very similar derivations can be applied to the models mentioned above.

## 2.2 Value of Information

Assume some CF technique that produces an explicit probabilistic model of the domain, giving rise to distributions over ratings for a specific user-product pair based on attributes of the user and product in question. For example, MCVQ, described in the next section, can be seen as producing a distribution over types for each product, a distribution over user attitudes towards products of each type, and a distribution over the ratings of product $j$ by user $i$ conditioned on their respective types and attitudes. A naive Bayes model [2] similarly produces a distribution over user "types" conditioned on a user's ratings, and a posterior over ratings given this distribution over types. For simplicity, we assume that the system can make recommendations only for a single product, and that the utility of any recommendation is given by its actual rating.[1] Thus the recommendation with highest expected utility is that product with highest mean rating. We define the *value* of the belief state $P_{\mathbf{r}_\kappa}$ (over ratings $R_{ij}$) to be

$$V(P_{\mathbf{r}_\kappa}) = \max_{j \in \overline{\kappa}} \sum_r r \cdot P_{\mathbf{r}_\kappa}(R_{ij} = r)$$

If we ask user $i$ to rate product $q \in \overline{\kappa}$ and get response $r_q$, our new posterior is $P_{\mathbf{r}_\kappa}^{r_q} = P_{\mathbf{r}_\kappa}(\cdot | R_{iq} = r_q)$, with value:

$$V(P_{\mathbf{r}_\kappa}^{r_q}) = \max_{j \in \overline{\kappa} \setminus \{q\}} \sum_r r \cdot P_{\mathbf{r}_\kappa}^{r_q}(R_{ij} = r)$$

The (myopic) *expected value of information* associated with query $q$ is the expected *improvement* in decision quality one obtains after asking $q$:

$$EVOI(q, P_{\mathbf{r}_\kappa}) = \sum_{r_q} \left( P_{\mathbf{r}_\kappa}(R_{iq} = r_q) V(P_{\mathbf{r}_\kappa}^{r_q}) \right) - V(P_{\mathbf{r}_\kappa})$$
(2)

In the myopic EVOI approach to active CF we ask the query whose EVOI is maximal (until some "query cost" threshold is reached).

It is important to note that this myopic approximation to true expected value of information can be led astray. For instance, if two queries could lead to a dramatic shift in our ratings prediction for a user, but neither query individually has any effect, myopic EVOI will be unable to discover this potentially valuable pair of queries.[2]

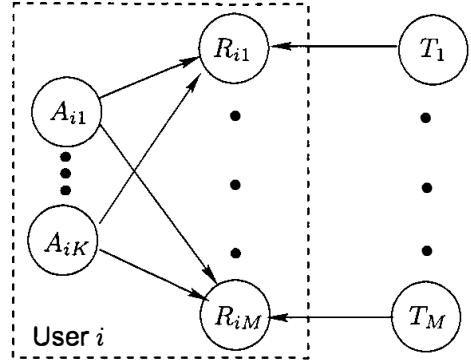

Figure 1: Graphical model for the MCVQ model. Circles denote random variables and the dashed rectangle shows the plate (i.e., repetitions) over the data (users).

## 3 Active CF in the MCVQ Model

In this section, we describe the MCVQ model developed by Ross and Zemel [10] and detail the derivation of the EVOI equations for this model. We emphasize, however, that EVOI can be applied to any explicit probabilistic model for CF. To illustrate this fact, we will include empirical results on the use of EVOI in a simple naive Bayes model.

### 3.1 CF with Multiple-Cause Vector Quantization

MCVQ is a probabilistic model for unsupervised learning which is particularly relevant to CF. The key assumption is that dimensions of the data can be separated into several disjoint subsets, or *multiple causes*, which take on values independently of each other. We also assume each cause is a *vector quantizer*, i.e., a multinomial with a small number of discrete states. Given a set of training examples, the MCVQ model learns the association of data dimensions with causes, as well as the states of each VQ. In the context of CF, the causes could correspond to *types* of items or products, and the states of a particular type could correspond to a user's *attitudes*, or rating profiles that a user can adopt towards items of the given type. In a music rating database, for example, each piece of music could be considered as a mixture of types or genres, and a user can be described as a mixture of attitudes towards each type, where each attitude implies a particular distribution over ratings for each piece of that type. In different terms, a particular user can be described as a *composite sketch*: a selection of the attitudes towards each type.

The notation and basic equations of MCVQ are as follows. Each item $j$ is one of $K$ types, or VQs: $P(T_j = k), k \in \{1, ..., K\}$. Corresponding to each type $k$ there are $L$ different attitudes that a user can adopt: $P(A_{ik} = l), l \in \{1, .., L\}$. Distributions over ratings can be estimated given these two quantities: $\theta^r_{jkl} \equiv P(R_{ij} = r | T_j = k, A_{ik} = l)$,

---
[1] Other decision criteria can be used with suitable modification to the value equations.

[2] Solutions to this problem include using multistage lookahead, or more accurately, modeling the entire interactive process as a sequential decision problem, much like the process of preference elicitation described in [1]. We leave the study of these more computationally demanding approaches to future work.

and the posterior over an item's rating is:

$$P_{\mathbf{r}_\kappa}(R_{ij} = r) = \sum_k P(T_j = k) \sum_l P(A_{ik} = l|\mathbf{r}_\kappa)\theta^r_{jkl}$$

offline from the large ratings database: $P(T_j = k)$, $\theta^r_{jkl}$, and $P(A_{ik} = l)$. The only online computation in the model entails updating the attitude distributions as more item ratings are observed:

$$P_{\mathbf{r}_\kappa}(A_{ik} = l) \equiv P(A_{ik} = l|\mathbf{r}_\kappa)$$
$$= \alpha \prod_{j \in \kappa} [\sum_{k' \neq k} P(T_j = k') \sum_{l'} P(A_{ik'} = l')\theta^{r_h}_{jk'l'}$$
$$+ P(T_j = k)\theta^{r_j}_{hkl}]P(A_{ik} = l)$$

where $\alpha$ is a normalizing constant.[3] A variational EM algorithm is used to learn the model parameters, and infer hidden variables (attitude distributions) given observations. Details of learning and inference in the model, as well as a discussion of its advantages, can be found in [10].

An example application of MCVQ to CF involves the EachMovie dataset, the database used for the experiments described in this paper. The dataset contains ratings, on a scale from 1 to 6, of a set of 1649 movies, by 74,424 users. Many of the movies are rated by very few users, and many users rate very few movies. We reduced the full dataset to one that includes all users who rated at least 75 movies and all movies rated by at least 126 users, leaving a total of 1003 movies and 5831 users. We create several random splits of this dataset into training and test sets, with 1000 users in each test set, leaving 4831 users in the corresponding training set. The parameters of an MCVQ model with 12 VQs and 4 components per VQ are learned on a training dataset.[4] An example of the results, after 15 iterations of EM, is shown in Figure 2.

### 3.2 EVOI in the MCVQ Model

The computations involved in calculating myopic EVOI in the MCVQ model are reasonably straightforward. We develop these in this section, but emphasize that the application of EVOI to other CF models would proceed in an analogous fashion. We assume $M$ products, $K$ types (or VQs), $L$ user attitudes toward products of a specific type (or components), and rating set $\{1, ..., \rho\}$. We assume a trained MCVQ model with parameters: $P(T_j = k)$, for $j \leq M, k \leq K$; $P(A_{ik} = l)$, for $k \leq K, l \leq L$; and $P(R_{ij} = r|T_j = k, A_{ik} = l) = \theta^r_{jkl}$, for $j \leq M, k \leq$ $K, l \leq L, r \leq \rho$. Note that the parameters $\theta^r_{jkl}$ are independent of the user $i$, and that $R_{ij}$ is independent of $A_{ik'}$ given $T_j = k$, for any $k' \neq k$.

---

[3]In order to reduce the number of parameters in the model, ratings are treated as continuous variables, so each state $l$ of type $k$ has a mean $\bar{\theta}_{jkl}$ and variance $\sigma^2_{jkl}$ in its rating predictions for item $j$. These are converted into multinomial distributions over ratings through binning and normalization.

[4]We use this same configuration of the MCVQ model throughout the paper. The model performance varies somewhat for different numbers of VQs and components per VQ, but this variation is not the central focus of this paper.

| VQ 2 | VQ 6 |
|---|---|
| The Shawshank Redemption 5.5 (5) | The Godfather 5.8 (6) |
| Taxi Driver 5.3 (6) | Pulp Fiction 5.7 (5) |
| Dead Man Walking 5.1 (-) | Get Shorty 5.2 (-) |
| Billy Madison 3.2 (-) | Sound of Music 2.9 (2) |
| Clerks 3.0 (4) | Lawrence of Arabia 2.6 (3) |
| Forrest Gump 2.7 (2) | Mary Poppins 2.4 (1) |
| Sling Blade 5.4 (5) | Mary Poppins 5.3 (5) |
| One Flew ... Cuckoo's Nest 5.3 (6) | The Wrong Trousers 5.2 (6) |
| Dr. Strangelove 5.2 (5) | Willy Wonka 5.0 (6) |
| The Beverly Hillbillies 2.0 (-) | Married to the Mob (3.3) 4 |
| Canadian Bacon 1.9 (4) | Pulp Fiction 3.2 (2) |
| Mrs. Doubtfire 1.7 (-) | GoodFellas 2.9 (2) |

Figure 2: The MCVQ representation of two test users in the EachMovie dataset. The 3 most conspicuously high-rated (bold) and low-rated movies by the most dominant attitudes (states) of 2 of the 12 VQs are shown (conspicuousness is the deviation from mean rating for a given movie). Each state's prediction, $\bar{\theta}_{jkl}$, can be compared to the user's true rating (in parentheses). Note the intuitive decomposition of movies into separate VQs, and that different states within a VQ may predict different rating patterns for the same movies.

Expected value of information can be computed in the fashion described above in the MCVQ model. The specifics of the MCVQ model dictate only how to update ratings distributions given a response to a query. Assume a user $i$ has provided response $r_q$ to query $q$. We then compute the posterior for any $j \in \overline{\kappa} \setminus q$:

$$P^{r_q}_{\mathbf{r}_\kappa} = P(R_{ij} = r|R_{iq} = r_q, \mathbf{r}_\kappa)$$
$$= \sum_{k \leq K} P(T_j = k) \sum_{l \leq L} \theta^r_{jkl} P_{\mathbf{r}_\kappa}(A_{ik} = l|R_{iq} = r_q)$$
$$= \sum_{k \leq K} P(T_j = k) \sum_{l \leq L} \theta^r_{jkl} \alpha [\sum_{k' \neq k} \{P(T_q = k')$$
$$\sum_{l' \leq L} P_{\mathbf{r}_\kappa}(A_{ik'} = l')\theta^{r_q}_{qk'l'}\} + P(T_q = k)\theta^{r_q}_{qkl}] P_{\mathbf{r}_\kappa}(A_{ik} = l)$$

Given these posterior calculations we can compute EVOI of any query using Eq. 2 above.

We evaluate the efficacy of this approach empirically by examining the change in *model loss* for the MCVQ model as we update ratings based on responses to queries. Model loss is defined as the difference between the user's actual utility (rating) for the best item we could have recommended and the actual utility for the item recommended by the model. The model recommendation is the item with highest mean rating (i.e., the item *predicted* to be best). xxxxxxxx We fix $|\kappa|$, the number of observed ratings, and randomly select items to be observed for each test user, holding out ratings of other items by this user. We then compute the model loss for those observations by subtracting the user's true rating of the model's highest ranked held-out item from the user's rating of her highest-ranked held-out item. We evaluate the *change* in model loss due to a query $q$ by observing the rating of item $q$, updating the model, and comparing the loss of the prior and posterior model (where the posterior model loss is defined over the reduced set of held-out items).[5]

We compare the change in model loss using the query with maximum EVOI with that obtained using two other query strategies. First, we test randomly generated queries, using these as a baseline.[6] We also compare EVOI-based querying with the entropy-based approach proposed in [9] as a method for populating the rating space for a new user. In this method, new users are asked about movies whose current empirical rating distribution (based on existing user ratings) has the greatest entropy.

Figure 3(a) shows our results using the learned MCVQ model. It demonstrates that selecting the held-out item to query based on EVOI leads to significantly greater improvements in model loss than either the random selection strategy or the entropy-based strategy, particularly for small values of $|\kappa|$. This dependence on $|\kappa|$ conforms with the intuition that the value of information decreases with increased user knowledge, as the posterior over ratings stabilizes. In fact, we could use a threshold on EVOI as a form of "query cost", so that if the maximum EVOI value does not exceed the threshold, the system stops querying, instead making a recommendation.[7] It is interesting to note that the entropy-based method is virtually indistinguishable from the random method with respect to model loss improvement.

Figure 3(b) shows a similar comparison of EVOI-based to random and entrop-based querying for a naive Bayes CF model learned on the same data. The model used is straightforward; the results shown are obtained with a model containing 40 components, and multinomial rating predictions. We do not provide a derivation, but note that the EVOI equations for this model are similar in flavor to those for MCVQ (though simpler). We see that EVOI again offers significant improvements, relative to both random and entropy-based query selection, in decision quality. This illustrates the general applicability of our active CF approach. The greater improvement in decision quality obtained by MCVQ is likely due to the greater flexibility in that model; the latent factor distribution is a product of several mixtures, rather than the single mixture in the naive Bayes model. The value of total loss is greater in MCVQ than naive Bayes when only one or two ratings are observed; the loss for MCVQ drops below naive Bayes with more observed ratings. The ability of the MCVQ model to tailor its posterior over latent factors to a user is reflected both in this drop in model loss, and the difference in the loss obtained by EVOI querying and the other strategies depicted in Figure 3.

## 4 Bounding Mean Rating Change

The straightforward computation of the EVOI of a query $q$ in the MCVQ model requires $O(\rho M)$ posterior computations. Since each unrated product is a potential query, determining the query with maximum EVOI requires $O(\rho M^2)$ posterior calculations. Since this process must be engaged online, while interacting with the user, this approach to active CF is unlikely to be feasible.

Fortunately, we can reduce the number of posterior calculations by bounding the impact a specific rating associated with product $q$ can have on the mean rating of product $j$. We do this in a user independent fashion, allowing the computation of these bounds offline (e.g., at the same time a new model is being learned with a new batch of data). As before, we assume a learned MCVQ model.

We first bound the difference in the posterior probability of a rating $P_{\mathbf{r}_\kappa}^{r_q}(R_{ij} = r) = P(R_{ij} = r | R_{iq} = r_q, \mathbf{r}_\kappa)$ given response $r_q$ to query $q$ and the prior $P_{\mathbf{r}_\kappa}(R_{ij} = r)$:

$$P_{\mathbf{r}_\kappa}^{r_q}(R_{ij} = r) - P_{\mathbf{r}_\kappa}(R_{ij} = r)$$
$$= \sum_{k \leq K} P(T_j = k) \sum_{l \leq L} [P_{\mathbf{r}_\kappa}^{r_q}(A_{ik} = l) - P_{\mathbf{r}_\kappa}(A_{ik} = l)]\theta_{jkl}^r$$
$$\leq \sum_{k \leq K} P(T_j = k) \sum_{l \leq L} \Delta_{kl}^{qr_q} \theta_{jkl}^r$$

where $\Delta_{kl}^{qr_q}$ is a bound on the term $|P_{\mathbf{r}_\kappa}^{r_q}(A_{ik} = l) - P_{\mathbf{r}_\kappa}(A_{ik} = l)|$ for any user $i$. Notice that the impact of a query rating on our predictions for a user $i$ is solely mediated by its impact on the user's attitude vector.

A bound can be derived by assuming a "worst case" distribution over user attitudes, one that maximizes the impact of the query rating on the target product rating. We omit details of the derivation, but note that the maximum decrease in $P^{r_q}(A_{ik} = l)$ is bounded by:

$$\frac{F + H_l}{F + pH_l + (1-p)H_{l_x}} p - p \qquad (3)$$

where

$$F = \sum_{k' \neq k} P(T_q = k') \sum_{l' \leq L} P_{\mathbf{r}_\kappa}(A_{ik'} = l')\theta_{qk'l'}^{r_q}$$
$$H_{l'} = P(T_q = k)\theta_{qkl'}^{r_q}$$
$$p = \frac{H_l + F - \sqrt{FH_l + F^2 + FH_{l_x} + H_{l_x}H_l}}{H_l - H_{l_x}}$$

---

[5]We must emphasize that as the number of known ratings was increased, held out ratings for a predetermined sequence of items were added to the set of known ratings for each user, and not the true rating for the previously queried item. In this sense, the results do not reflect the sequential nature of the interaction in which a user would be engaged. This procedure is only an approximation to the interactive rating entry process, and is adopted simply to show how different query strategies, and different CF models, behave given the *same* input data at any given stage.

[6]Queries are restricted to held-out items, since these are the only queries for which can can obtain actual "responses."

[7]If we had pursued this strategy, we expect that the improvement per query would be considerably greater than is shown, because for many users the maximum EVOI was negligible.

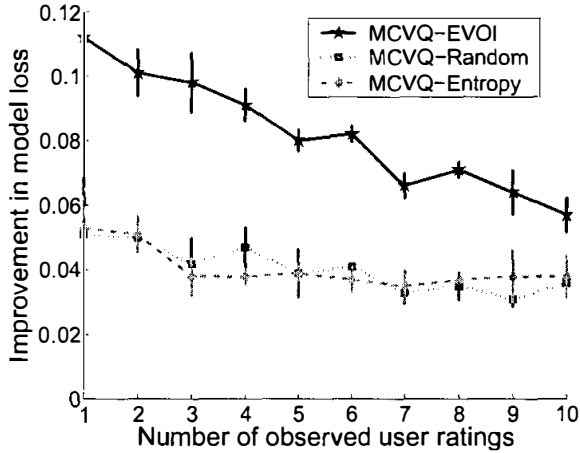
(a) Results using MCVQ

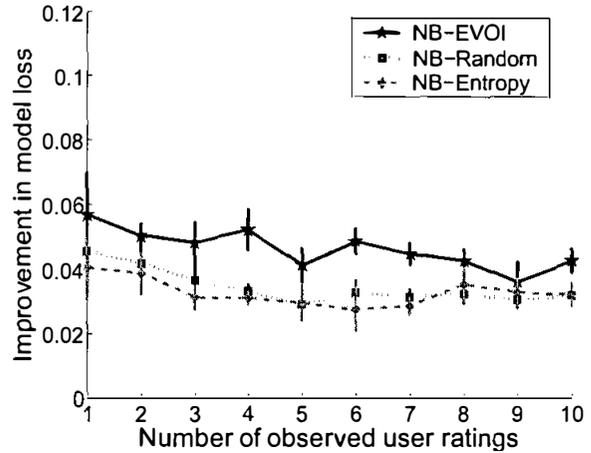
(b) Results using Naive Bayes

Figure 3: The average improvement in model loss (difference between actual and predicted utility) for (a) MCVQ and (b) naive Bayes for varying number of observed ratings of test users. Three query strategies are show: EVOI-based, entropy-based, and random. Recall that the improvement is for ratings that lie with in the range 1–6. The plots average the improvement across the set of 1000 test users. Each data point is an average of 5 runs (with standard error bars), with different training and test sets, and a random selection of observed ratings on each run. These sets and ratings were identical for both models and all three strategies.

An analogous expression exists for the maximum increase. Thus we set the bound $\Delta_{kl}^{qr_q}$ to the maximum (in absolute terms) of the maximum increase/decrease.

We can derive an analytic expression that bounds mean rating change using the $\Delta_{kl}^{qr}$ terms, but a much tighter bound is achievable by explicitly modeling the prior of $R_{ij}$ and finding a worst-case distribution $P(R_{ij})$ that maximizes $V_j^{qr_q} - V_j$ (the minimum has the same absolute value). This can be accomplished with a very compact linear program (LP). We use variables $p_1, \ldots, p_\rho$ denoting the prior $P(R_{ij})$ of each rating $r \leq \rho$; $q_1, \ldots, q_\rho$, denoting the posterior over $R_{ij}$; and $\delta_{kl}$ for each type-attitude pair $k, l$, denoting the actual change in $P(A_{ik} = l)$ in response to the query. We impose standard simplex contraints on the variables $p_r$ and $q_r$. We also impose the bounds $-\Delta_{kl}^{qr} \leq \delta_{kl} \leq \Delta_{kl}^{qr}$. Finally, we relate the change in attitude distributions to the change in rating distributions by imposing the following equality constraint for each $r \leq \rho$:

$$q_r - p_r = \sum_k P(T_j = k) \sum_l \theta_{kl}^r \delta_{kl}$$

Maximizing the objective function $\sum_r r \cdot (q_r - p_r)$ subject to these constraints bounds the change in mean rating. We set $\Delta_j^{qr_q}$ to the objective value obtained by solving this LP.

The LP for each $\Delta_j^{qr_q}$ is very compact, with $KL + 2\rho$ variables, and $2KL + 5\rho + 2$ constraints. We do note that this bound can also be produced using a simple iterative algorithm with complexity $O(KL\rho)$ (we omit details). In practice, however, it appears that the direct LP formulation can be solved very effectively.

With this procedure in place, we can compute the set of terms $\Delta_j^{qr}$ for each product $j$, query (product) $q$, and query response (rating) $r$. While this computation is significant, again we emphasize that it is performed offline given a stable learned model, and is user-independent. These terms can be used to prune the number of posterior computations needed to compute the query with maximum EVOI. Let $j^*$ be the product with highest mean rating for user $i$. For a specific query $q$, we can forego the computation of the posterior $P_{r_\kappa}(R_{ij} = r | R_{iq} = r_q)$ (for each response $r_q$) if our bounds preclude the possibility of the mean of $R_{ij}$ becoming higher than that of $R_{ij^*}$. More precisely, if:

$$V_{j^*} - \sum_r P_{r_\kappa}(R_{iq} = r)\Delta_{j^*}^{qr} \geq V_j + \sum_r P_{r_\kappa}(R_{iq} = r)\Delta_j^{qr} \quad (4)$$

then we need not compute the posterior over $R_{ij}$ when computing EVOI of $q$. This can offer significant pruning.

We empirically evaluate the amount of pruning obtained by this approach using a procedure similar to the experiment presented in the previous section. We use the same trained MCVQ model as above. We observe $|\kappa|$ ratings of a given test user $i$, and update the attitude distributions and posterior over $r_{\overline{\kappa}}$, the ratings of unobserved items. For each possible query item $q$ and target $j \in \overline{\kappa}$, we compute $\Delta_j^{qr}$, as well as $\Delta_{j^*}^{qr}$ for each $q$. For each movie $j$ we can then apply Eq. 4 to determine if that movie cannot possibly obtain a higher rating than the model's current top-rated movie after query $q$. The number of movies satisfying this inequality describes the degree of pruning in posterior computations.

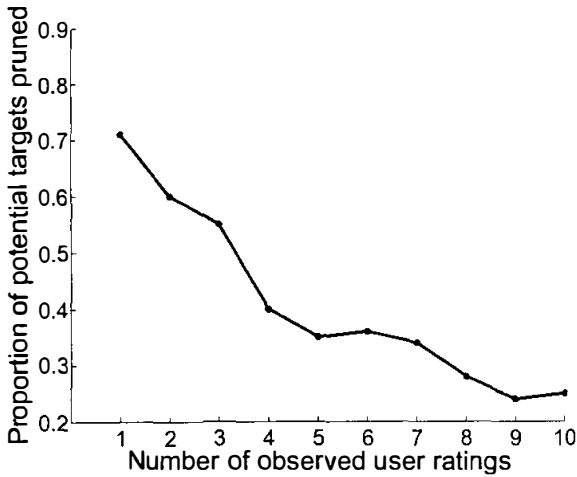

Figure 4: The proportion of unobserved items for which posterior distributions need not be computed is plotted for varying number of observed ratings of test users. As before, each datapoint an average of 5 runs for the model, with a random selection of observed ratings for the test users on each run.

Figure 4 plots the pruning of potential targets as the ratio of number of unobserved movies not satisfying Eq. 4 to potential targets $(M - |\kappa| - 1)$. The figure shows a large degree of pruning at the early stages of interaction with the user, but is fairly substantial throughout the interaction period. This implies that many items do not have the potential of ever surpassing the estimated utility of the model's top-ranked item, and substantial computational savings can be obtained by identifying these based on computations that can occur primarily offline.

While pruning is significant, the fact that the algorithm uses no online, user-specific information to prune suggests that more aggressive pruning is possible. However, our main aim is to ensure that all significant computation is effected offline; hence we do not want to recompute bounds online using the current user's attitude distrubution.

We can get the best of both worlds by "clustering" user distributions offline, deriving suitable $\Delta_j^{qr}$ terms for each such cluster. Given a particular subspace of the space of possible user distributions, we can apply the technique described above to compute the $\Delta_j^{qr}$ terms with the restriction that the user distribution lie somewhere in this subspace. For instance, we might impose a simple linear constraint that $P(A_{ik} = l) \leq 0.5$, and compute the greatest possible change terms $\Delta_j^{qr}$ subject to this constraint. Note that this LP is identical to the one above with one additional constraint. Generally, the maximum mean ratings changes will be smaller given such a restriction, allowing more aggresive pruning. Of course, we would need to partition the set of user distributions into some finite set of such "prototypes" (e.g., by imposing a collection of linear constraints that define each prototype) and derive a distinct set of $\Delta_j^{qr}$ terms for each such prototype. Note that the increase in computational cost will be proportional to the number of prototype clusters constructed, but this cost is borne entirely offline.

Online pruning proceeds exactly as above, with the exception that we first identify the cluster to which the current user's attitude distribution belongs (this can change as our posterior changes), and then apply Eq. 4 using the cluster-specific $\Delta$-terms. We are currently experimenting with this approach and will report on results in an extended version of this paper.

## 5 Prototype Queries

The bounds in the previous section restrict the number of posterior computations over target products for each query-rating pair to those that could possibly become optimal; this reduces the $O(\rho M^2)$ problem to $O(\rho M N)$, where $N \leq M$ is the expected number of targets for which posteriors must be computed. This depends on the degree of pruning possible for a specific problem, but as we've seen, $N$ appears to be considerably less than $M$ in practice.

We might also attempt to reduce the number of queries we need to consider: considering $Q < M$ queries reduces online posterior computations to $O(\rho Q N)$. In this section we describe a simple method for offline construction of a set of $Q$ *prototype queries*, with the property that the EVOI of any query $m \in M$ is within some bound $\varepsilon$ of some prototype query $q \in Q$. By restricting attention to queries in $Q$, we reduce online complexity further, but guarantee $\varepsilon$-optimal querying behavior.

Intuitively, the difference in the impact of two potential queries $q$ and $q'$ can be characterized by the difference in the type distributions of each query, and the difference in their rating parameters. For any product (i.e., potential query) $q$, define $v_q$ to be a vector of length $KL\rho$ with elements $\Pr(T_q = k)\theta_{qkl}^r$. For two queries $q$ and $q'$, the fundamental distinction between $q$ and $q'$ can be characterized by the $L_1$-distance $d(q, q') = ||v_q - v_{q'}||_1$ between these vectors.

The key fact to notice is that the difference in $\Delta_{kl}^{qr}$ and $\Delta_{kl}^{q'r}$ (for any $k, l, r$) is bounded by a continuous function $f$ of $d(q, q')$; that is, if $d(q, q') \leq \varepsilon$, then $|\Delta_{kl}^{qr} - \Delta_{kl}^{q'r}| \leq f(\varepsilon, k, l, r)$. We currently have some fairly crude bounds that are independent of all terms except $\varepsilon$, as well as a somewhat more reasonable approximation $f(\varepsilon, r) = 12\varepsilon/P(r)$, where $P(r)$ is the probability of receiving response $r$ under query $q$.[8] From this, we can bound the difference between the terms $\Delta_j^{qr}$ and $\Delta_j^{q'r}$ for each target $j$ with the same $f(\varepsilon)$. Finally, we obtain a bound on the dif-ference in the expected mean rating change in target $j$ due to query $q$ and query $q'$ via $\sum_r P(R_{iq} = r)f(\varepsilon, k, l, r)$. For example, we obtain

$$|V_j^q - V_j^{q'}| \leq 12\varepsilon$$

using $f(\varepsilon, r) = 12\varepsilon/P(r)$. Here $V_j^q$ denotes the expected value (mean rating) of product $j$ after receiving a response to query $j$.

---
[8] We expect that much tighter bounds than the ones we have derived currently are possible.

This suggests an obvious method for constructing query protoypes that reduce the number of queries one needs to consider to guarantee that a query is chosen that has approximately optimal myopic EVOI. Given a learned model, our aim is to construct a set of prototype queries $Q$ such that, for any product $j$, there exists a product $q \in Q$ such that $d(q, j) \leq \varepsilon$. This is a straightforward clustering task. If we restrict our attention to such a set and chose the query within $Q$ that has maximum EVOI, we can guarantee that we are acting $f(\varepsilon)$-optimally with respect to considering the full set of potential queries.

We have performed some preliminary experimentation with prototype queries. While there are certainly more sophisticated techniques for identifying prototypes, we used a simple greedy strategy to identify a set of queries with the property that no two queries in the set have distance $d(q, q') \leq \beta$. This distance is not the sole criterion for selecting prototypes; another important consideration is the likelihood that a particular query will be answerable by the user. We take this into account by ordering the movies in the EachMovie dataset according to the number of ratings they have received. A set of prototypes biased towards answerable queries can then be produced by considering each of the movies in this list in order, adding it to the set if its distance from any prototype is at least $\beta$.

We employed this simple approach to produce two sets of prototype queries: one in which roughly 60% of the total potential queries were pruned to leave 406 movies; and a second in which about 80% of the total queries were pruned to leave 215 possible queries. We then tested EVOI-based querying using these restricted set of queries to the EVOI using the full set of queries. A comparison of the reduction in model loss for both approaches is shown in Figure 5. (We use the same methodology as above, so the curve for the full query set duplicates that in Figure 3.)

We see that even using the naive prototype generation strategy described above, we obtain very good improvement in model-loss using EVOI-based querying when queries are severely restricted. At 60% pruning (leaving only 40% of possible movies available for querying), we see improvement in model-loss that tracks that of the unpruned query set reasonably closely. At the more aggressive 80% pruning level, we see less improvement in model-loss; but it still offers significantly better performance than the other query strategies (random querying is included in the figure for reference). These results are encouraging; even with very aggressive pruning—and the concomitant reduction in online computational cost—our active EVOI-based works extremely well. With more intelligent prototype generation techniques, we fully expect this approach to work even more effectively.

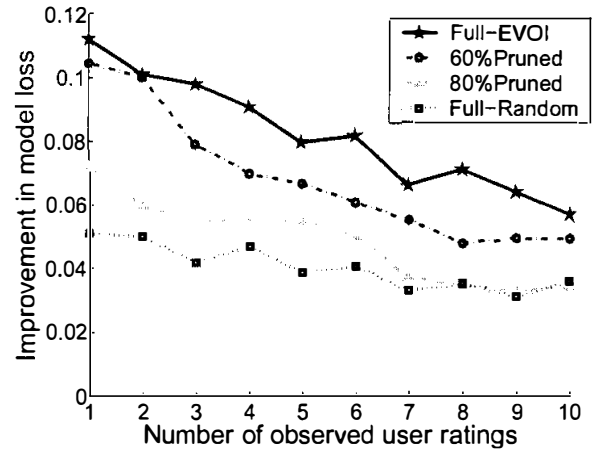

Figure 5: The normalized total improvement in model loss (difference between actual and predicted utility) for MCVQ for varying number of observed ratings of test users. Results from two different degrees of query prototyping (80 and 60% of all potential queries pruned away) are shown, with the MCVQ results shown in Figure 3 repeated for comparison.

## 6 Concluding Remarks

We have proposed an active approach to CF, based on a probabilistic model of user preference data. Our framework is quite general, considering the value of queries that could most improve the quality of the recommendations made, based on the model's predictions. We have shown that offline pre-computation of bounds on value of information, and of prototypes in query space, can be used to dramatically reduce the required online computation. We also have derived detailed bounds for a particular model, and empirically demonstrated the value of our active approach using this model. While off-line computations should also lead to considerable savings in other probabilistic models, we expect the savings in MCVQ to be greater due to the user-independent assignment of movies to types.

While active CF has been discussed in the literature, specific models based on the use of value of information have not been investigated. Pennock and Horvitz [8] have suggested the use of EVOI, but did not propose specific models nor address the computational difficulties associated with its use in CF. Other active strategies have been investigated in some depth in [9]. This work, in the context of the "new user problem," suggests methods for populating the rating space for a new user by actively querying them for specific movie ratings. However, the methods they consider, including asking for ratings of movies that have been rated the most by existing users, or entropy-based methods—i.e., asking about movies whose current empirical rating distribution based on existing users has the greatest entropy—bear little relation to our EVOI-based approach. However, as we have discussed above, properties such as how likely a movie is to be "rate-able" should play a role in active methods, hence there is certainly an opportunity to incorporate such methods with ours in a decision-theoretic manner.

Finally, we note that while the empirical results presented here are encouraging, these results are hampered by the nature of the dataset we used. The sparseness of the data makes statistical evaluation difficult; for example, the pool of test users shrinks considerably when we increase the number of known ratings beyond a few, so that evaluating EVOI-based querying becomes problematic. In addition, the static nature of the dataset means that we can only evaluate queries for items that the user actually rated. We are planning to develop an active collaborative filtering prototype system for user testing. This will allow us to overcome the problem with a fixed dataset, as we can query users by providing a video or music clip of a queried item. In addition we can use this system to directly compare different query strategies and CF models.

Other current directions of this work include improving the bounds, approximate pruning of targets, and further studies of prototyping of queries. In addition, we are examining alternative cost models for queries, including modeling the probability that a user can answer a given query. We are also exploring the use of this methodology to tackle different, though related tasks. For example, if the goal is not to recommend a movie, but to accurately predict whether a user will like a *specific* movie, active techniques can again be used to improve prediction quality. In such a scenario, we expect that pruning and prototyping will be much more effective. Finally, we are considering extending the myopic approach to examine multistage lookahead, and offline policy construction. An important challenge in this endeavor will be the development of computationally feasible strategies to solve for approximately optimal policies.

## Acknowledgements

We thank Scott Helmer for his work on the movie database and on exploring other probabilistic methods of CF. We also thank David Ross for his work on the multiple cause vector quantization algorithm. We thank the Compaq Equipment Corporation for making the EachMovie dataset available. This research was supported by grants from NSERC and IRIS. An earlier version of this paper (which excluded evaluation of naive Bayes models and query prototyping) appeared in Ninth International Workshop on Artificial Intelligence and Statistics (AI-Stats).